\documentstyle[floats,prl,aps]{revtex}
\begin{document}
\renewcommand{\thefootnote}{\fnsymbol{footnote}}
\draft
\title{\Large\bf 
Boundary two-parameter eight-state supersymmetric fermion model and
Bethe ansatz solution}

\author{ Anthony J. Bracken, Xiang-Yu Ge 
 \footnote {xg@maths.uq.edu.au},
Yao-Zhong Zhang \footnote {yzz@maths.uq.edu.au}}

\address{Department of Mathematics, University of Queensland,
		     Brisbane, Qld 4072, Australia}

\author{ Huan-Qiang Zhou }
\address{CCAST (World Laboratory), P.O.Box 8730, Beijing 100080, China\\ and
    Department of Physics, Chongqing University, Chongqing
    630044, China}
\maketitle

\vspace{10pt}

\begin{abstract}
The recently introduced two-parameter eight-state $U_q[gl(3|1)]$
supersymmetric fermion model is extended to include boundary
terms. Nine classes of boundary conditions are constructed,
all of which are shown to be integrable via the graded boundary quantum
inverse scattering method. The boundary systems are solved 
by using the coordinate Bethe ansatz and the Bethe ansatz
equations are given for all nine cases. 
\end{abstract}




\def\a{\alpha}
\def\b{\beta}
\def\d{\delta}
\def\e{\epsilon}
\def\g{\gamma}
\def\k{\kappa}
\def\l{\lambda}
\def\o{\omega}
\def\t{\theta}
\def\s{\sigma}
\def\D{\Delta}
\def\L{\Lambda}


\def\beq{\begin{equation}}
\def\eeq{\end{equation}}
\def\bea{\begin{eqnarray}}
\def\eea{\end{eqnarray}}
\def\ba{\begin{array}}
\def\ea{\end{array}}
\def\no{\nonumber}
\def\le{\langle}
\def\re{\rangle}
\def\lt{\left}
\def\rt{\right}

\newcommand{\sect}[1]{\setcounter{equation}{0}\section{#1}}
\renewcommand{\theequation}{\thesection.\arabic{equation}}
\newcommand{\reff}[1]{eq.~(\ref{#1})}

\vskip.3in

\sect{Introduction}

Low dimensional integrable quantum systems, with or without open
boundary conditions, which describe 
strongly correlated fermions \cite{Ess94} form an important class
of lattice integrable models, which have attracted much international
attention (see, e.g. \cite{Bar91,Bra95,Bar95,Asa96} and references
therein). Recently, trying to extend the existing two component
electron models to
multi-component cases, we proposed \cite{Gou97}
an eight-state integrable model and its two-parameter (or $q$-deformed)
version with Lie superalgebra $gl(3|1)$ and 
quantum superalgebra $U_q[gl(3|1)]$ symmetries, respectively. 
One of the features of these two models
is that they contain correlated single-particle and
pair hoppings, uncorrelated triple-particle hopping and two- and
three-particle on-site interactions. By eight-state, we mean that
at a given lattice site $j$ of the length $L$
there are eight possible fermionic states:
\bea
&&|0\re\,,~~~c_{j,1}^\dagger|0\re\,,~~~
  c_{j,2}^\dagger|0\re\,,~~~ c_{j,3}^\dagger|0\re,\no\\
&&c_{j,1}^\dagger c_{j,2}^\dagger|0\re\,,~~~ 
  c_{j,1}^\dagger c_{j,3}^\dagger|0\re\,,~~~ 
  c_{j,2}^\dagger c_{j,3}^\dagger|0\re\,,~~~ 
  c_{j,1}^\dagger c_{j,2}^\dagger c_{j,3}^\dagger|0\re\,,\label{states}
\eea
where $c_{j,\a}^\dagger$ ($c_{j,\a}$) denotes a fermionic creation
(annihilation) operator which creates (annihilates) a fermion
of species $\a=1,\;2,\;3$ at
site $j$; these operators satisfy the anti-commutation relations given by
$\{c_{i,\a}^\dagger, c_{j,\b}\}=\d_{ij}\d_{\a\b}$.

Recently we formulated in \cite{Bra97} a general and fully supersymmetric
version of the boundary inverse scattering method \cite{Skl88,Mez91}, 
and constructed
a large number of integrable boundary conditions \cite{Zha97} for various
models of strongly correlated fermions. 
In this paper, we continue our study of open boundary conditions
and consider the integrable eight-state fermion model with 
$U_q[gl(3|1)]$ symmetry. We present nine classes of boundary
conditions for this model, all of which are shown to be integrable
by the graded boundary QISM \cite{Bra97}.
We solve the boundary systems by using the coordinate Bethe
ansatz method and derive the Bethe ansatz equations for all nine cases.

\sect{Open Boundary Conditions}

We consider the following Hamiltonian with boundary terms
\beq
H=\sum _{j=1}^{L-1} H_{j,j+1}(g,\k) + H^{\rm boundary}_{lt} 
+H^{\rm boundary}_{rt},\label{h}
\eeq
where $H^{\rm boundary}_L,~H^{\rm boundary}_R$ are respectively left and right
boundary terms whose explicit forms are given below, and
$H_{j,j+1}$ is the Hamiltonian density of the two-parameter
eight-state supersymmetric fermion model \cite{Gou97}
\bea
H_{j,j+1}(g,\k)&=&-\sum_\a(c_{j,\a}^\dagger c_{j+1,\a}+{\rm h.c.})
  \exp\lt\{-\frac{1}{2}(\eta +\kappa) \sum_{\b(\neq\a)}n_{j+\theta(\b-\a),\b}
  -\frac {1}{2}(\eta-\kappa) \sum _{\b(\neq \a)}
  n_{j+1-\theta (\b-\a),\b}\rt.\no\\
   & &\lt.+\frac{\zeta}{2}
  \sum_{\b\neq\g(\neq\a)}(n_{j,\b}n_{j,\g}
  +n_{j+1,\b} n_{j+1,\g})\rt\}
-\frac{\sinh \kappa}{2\sinh \kappa(g+1)}
\sum_{\a\neq\b\neq\g}(c_{j,\a}^\dagger c_{j,\b}^\dagger 
  c_{j+1,\b}c_{j+1,\a}+{\rm h.c.})\no\\
 & &\exp\lt\{-(\frac{\xi}{2}-{\rm sign}(\g-2)\kappa)n_{j,\g}-(\frac {\xi}{2}+
  {\rm sign}(\g-2)\kappa)n_{j+1,\g}\rt\}\no\\
& &-\frac{2\cosh \kappa \sinh^2 \kappa}{\sinh \kappa(g+1)\sinh \kappa
(g+2)}\lt(c_{j,1}^\dagger c_{j,2}^\dagger 
  c_{j,3}^\dagger c_{j+1,3} c_{j+1,2} c_{j+1,1}+{\rm h.c.}\rt)\no\\
& & +e^{\kappa g}\; n_{j}+e^{-\kappa g}\;
  n_{j+1}-\frac{\sinh \kappa}{2\sinh \kappa (g+1)}\sum_{\a\neq\b}
  (n_{j,\a}n_{j,\b}+n_{j+1,\a}n_{j+1,\b})\no\\
& &+\frac{2\cosh \kappa (g+1)\sinh^2 \kappa}{\sinh \kappa (g+1)\sinh
  \kappa (g+2)}(n_{j,1}n_{j,2}n_{j,3}+n_{j+1,1}n_{j+1,2}
  n_{j+1,3}),\label{hamiltonian}
\eea
where $g,~\k$ are two free parameters, $n_j=n_{j,1}+n_{j,2}+n_{j,3}$ with
$n_{j,\a}=c_{i,\a}^\dagger c_{j,\a}$ being the number operator
for the electron of species $\a$ at site $j$,
$\t(\b-\a)$ is a step function of $(\b-\a)$ and
\beq
\eta=-\ln\frac{\sinh \k g}{\sinh\k(g+1)},~~~~\zeta=\frac{1}{2}
  \ln\frac{\sinh^2\k(g+1)}{\sinh\k g\,\sinh\k(g+2)},~~~~
  \xi=-\ln\frac{\sinh\k g}{\sinh\k(g+2)}.
\eeq

As is shown in \cite{Gou97}, the symmetry algebra underlying 
(\ref{hamiltonian}) is quantum superalgebra $U_q[gl(3|1)]$. The
parameter $\k$ is related to the deformed parameter $q$ by
$q=e^\k$.

We propose the following nine classes of boundary conditions:
\bea
{\rm Case ~(i)}:~~H^{\rm boundary}_{lt}&=&\frac {\sinh\kappa g}
  {\sinh\frac{\xi^I_-}{2}\kappa}
\lt( e^{-\frac{\xi^I_-}{2}\kappa}n_1-
  \frac {\sinh\kappa}{\sinh(1+\frac{\xi^I_-}{2})\kappa}
( n_{1,1}n_{1,2}+n_{1,2}n_{1,3}+n_{1,1}n_{1,3})\rt. \no\\
& &  \lt.+\frac {2\sinh^2\kappa \cosh(1+\frac{\xi^I_-}{2})\kappa}
  {\sinh(1+\frac{\xi^I _-}{2})\kappa\sinh
  (2+\frac{\xi^I _-}
  {2})\kappa}n_{1,1} n_{1,2} n_{1,3}\rt) ,\no\\
H^{\rm boundary}_{rt}&=&
\frac{\sinh\kappa g}
  {\sinh\frac{\xi^I_+}{2}\k}
\lt(e^{\frac{\xi^I_+}{2}\kappa}n_L-
  \frac{\sinh\kappa}{\sinh(1+\frac{\xi^I_+}{2})\kappa}
  (n_{L,1}n_{L,2}+n_{L,2}n_{L,3}+n_{L,1}n_{L,3})\rt.\no\\
& & \lt. +\frac {2\sinh^2\kappa \cosh(1+\frac{\xi^I_+}{2})\kappa}
  {\sinh(1+\frac{\xi^I _+}{2})\kappa\sinh
  (2+\frac{\xi^I _+}{2})\kappa}n_{L,1} n_{L,2} n_{L,3}\rt) 
  ;\label{boundary11}\\
{\rm Case ~(ii)}:~~H^{\rm boundary}_{lt}&=&\frac {\sinh\kappa g}
  {\sinh\frac{\xi^{II}_-}{2}\kappa}\lt(e^{-\frac{\xi^{II}_-}{2}\kappa}
  (n_{1,2}+n_{1,3})-\frac{\sinh\kappa}{\sinh(1+\frac{\xi^{II}_-}
  {2})\kappa}n_{1,2}n_{1,3}\rt),\no\\
H^{\rm boundary}_{rt}&=&\frac {\sinh\kappa g}
  {\sinh\frac{\xi^{II}_+}{2}\kappa}\lt(e^{\frac{\xi^{II}_+}{2}\kappa}
  (n_{L,2}+n_{L,3})- \frac{\sinh\kappa}{\sinh(1+\frac{\xi^{II}_+}
  {2})\kappa}n_{L,2}n_{L,3}\rt)
  ;\label{boundary22}\\
{\rm Case ~(iii)}:~~H^{\rm boundary}_{lt}&=&\frac{\sinh\kappa g}{\sinh
   \frac{\xi^{III}_-}{2}\kappa}e^{-\frac{\xi^{III}_-}{2}\kappa}n_{1,3},\no\\
H^{\rm boundary}_{rt}&=&\frac{\sinh\kappa g}{\sinh
   \frac{\xi^{III}_+}{2}\kappa}e^{\frac{\xi^{III}_+}{2}\kappa}n_{L,3}
;\label{boundary33}\\
{\rm Case ~(iv)}:~~H^{\rm boundary}_{lt}&=&
\frac {\sinh\kappa g}
  {\sinh\frac{\xi^I_-}{2}\kappa}
\lt(e^{-\frac{\xi^I_-}{2}\kappa}n_1-
  \frac {\sinh\kappa}{\sinh(1+\frac{\xi^I_-}{2})\kappa}
( n_{1,1}n_{1,2}+n_{1,2}n_{1,3}+n_{1,1}n_{1,3})\rt.\no\\
& & \lt. +\frac {2\sinh^2\kappa \cosh(1+\frac{\xi^I_-}{2})\kappa}
  {\sinh(1+\frac{\xi^I _-}{2})\kappa\sinh
  (2+\frac{\xi^I _-}
  {2})\kappa}n_{1,1} n_{1,2} n_{1,3}\rt) ,\no\\
H^{\rm boundary}_{rt}&=&\frac {\sinh\kappa g}
  {\sinh\frac{\xi^{II}_+}{2}\kappa}\lt(e^{\frac{\xi^{II}_+}{2}\kappa}
  (n_{L,2}+n_{L,3})- \frac{\sinh\kappa}{\sinh(1+\frac{\xi^{II}_+}
  {2})\kappa}n_{L,2}n_{L,3}\rt)
  ;\label{boundary12}\\
{\rm Case ~(v)}:~~H^{\rm boundary}_{lt}&=&\frac {\sinh\kappa g}
  {\sinh\frac{\xi^{II}_-}{2}\kappa}\lt(e^{-\frac{\xi^{II}_-}{2}\kappa}
  (n_{1,2}+n_{1,3})-\frac{\sinh\kappa}{\sinh(1+\frac{\xi^{II}_-}
  {2})\kappa}n_{1,2}n_{1,3}\rt),\no\\
H^{\rm boundary}_{rt}&=&
\frac{\sinh\kappa g}
  {\sinh\frac{\xi^I_+}{2}\k}
\lt(e^{\frac{\xi^I_+}{2}\kappa}n_L-
  \frac{\sinh\kappa}{\sinh(1+\frac{\xi^I_+}{2})\kappa}
  (n_{L,1}n_{L,2}+n_{L,2}n_{L,3}+n_{L,1}n_{L,3})\rt.\no\\
& & \lt. +\frac {2\sinh^2\kappa \cosh(1+\frac{\xi^I_+}{2})\kappa}
  {\sinh(1+\frac{\xi^I _+}{2})\kappa\sinh
  (2+\frac{\xi^I _+}{2})\kappa}n_{L,1} n_{L,2} n_{L,3}\rt) 
  ;\label{boundary21}\\
{\rm Case ~(vi)}:~~H^{\rm boundary}_{lt}&=&\frac {\sinh\kappa g}
  {\sinh\frac{\xi^I_-}{2}\kappa}
\lt(e^{-\frac{\xi^I_-}{2}\kappa}n_1-
  \frac {\sinh\kappa}{\sinh(1+\frac{\xi^I_-}{2})\kappa}
( n_{1,1}n_{1,2}+n_{1,2}n_{1,3}+n_{1,1}n_{1,3})\rt.\no\\
& & \lt. +\frac {2\sinh^2\kappa \cosh(1+\frac{\xi^I_-}{2})\kappa}
  {\sinh(1+\frac{\xi^I _-}{2})\kappa\sinh
  (2+\frac{\xi^I _-}
  {2})\kappa}n_{1,1} n_{1,2} n_{1,3}\rt) ,\no\\
H^{\rm boundary}_{rt}&=&\frac{\sinh\kappa g}{\sinh
   \frac{\xi^{III}_+}{2}\kappa}e^{\frac{\xi^{III}_+}{2}\kappa}n_{L,3}
;\label{boundary13}\\
{\rm Case ~(vii)}:~~H^{\rm boundary}_{lt}&=&\frac{\sinh\kappa g}{\sinh
   \frac{\xi^{III}_-}{2}\kappa}e^{-\frac{\xi^{III}_-}{2}\kappa}n_{1,3},\no\\
H^{\rm boundary}_{rt}&=&
\frac{\sinh\kappa g}
  {\sinh\frac{\xi^I_+}{2}\k}
\lt(e^{\frac{\xi^I_+}{2}\kappa}n_L-
  \frac{\sinh\kappa}{\sinh(1+\frac{\xi^I_+}{2})\kappa}
  (n_{L,1}n_{L,2}+n_{L,2}n_{L,3}+n_{L,1}n_{L,3})\rt.\no\\
& & \lt. +\frac {2\sinh^2\kappa \cosh(1+\frac{\xi^I_+}{2})\kappa}
  {\sinh(1+\frac{\xi^I _+}{2})\kappa\sinh
  (2+\frac{\xi^I _+}{2})\kappa}n_{L,1} n_{L,2} n_{L,3}\rt) 
  ;\label{boundary31}\\
{\rm Case ~(viii)}:~~H^{\rm boundary}_{lt}&=&\frac {\sinh\kappa g}
  {\sinh\frac{\xi^{II}_-}{2}\kappa}\lt(e^{-\frac{\xi^{II}_-}{2}\kappa}
  (n_{1,2}+n_{1,3})-\frac{\sinh\kappa}{\sinh(1+\frac{\xi^{II}_-}
  {2})\kappa}n_{1,2}n_{1,3}\rt),\no\\
H^{\rm boundary}_{rt}&=&\frac{\sinh\kappa g}{\sinh
   \frac{\xi^{III}_+}{2}\kappa}e^{\frac{\xi^{III}_+}{2}\kappa}n_{L,3}
;\label{boundary23}\\
{\rm Case ~(ix)}:~~H^{\rm boundary}_{lt}&=&\frac{\sinh\kappa g}{\sinh
   \frac{\xi^{III}_-}{2}\kappa}e^{-\frac{\xi^{III}_-}{2}\kappa}n_{1,3},\no\\
H^{\rm boundary}_{rt}&=&\frac {\sinh\kappa g}
  {\sinh\frac{\xi^{II}_+}{2}\kappa}\lt(e^{\frac{\xi^{II}_+}{2}\kappa}
  (n_{L,2}+n_{L,3})- \frac{\sinh\kappa}{\sinh(1+\frac{\xi^{II}_+}
  {2})\kappa}n_{L,2}n_{L,3}\rt)
  ;\label{boundary32}
\eea
where $\xi^{a}_{\pm}(a=I,II,III)$ are some parameters describing the
boundary effects. As will be shown in next section, all nine classes
of boundary conditions are integrable.

\sect{Boundary K-matrices and Quantum Integrability}

Quantum integrability of the boundary conditions (\ref{boundary11}--
\ref{boundary32}) can be established by means of 
the (graded) boundary QISM recently formulated in \cite{Bra97}. 
 We first search for boundary K-matrices which
satisfy the graded reflection equations:
\bea
&&R_{12}(u_1-u_2)\stackrel {1}{K}_-(u_1) R_{21}(u_1+u_2)
  \stackrel {2}{K}_-(u_2)
=  \stackrel {2}{K}_-(u_2) R_{12}(u_1+u_2)
  \stackrel {1}{K}_-(u_1) R_{21}(u_1-u_2),  \label{reflection1}\\
&&R_{21}^{st_1 ist_2}(-u_1+u_2)\stackrel {1}{K_+^{st_1}}
  (u_1) R_{12}(-u_1-u_2+4)
  \stackrel {2}{K_+^{ist_2}}(u_2)\no\\
&&~~~~~~~~~~~~~~~~~~~~~~~~~=\stackrel {2}{K_+^{ist_2}}(u_2) R_{21}(-u_1-u_2+4)
  \stackrel {1}{K_+^{st_1}}(u_1) R_{12}^{st_1 ist_2}(-u_1+u_2)
  ,\label{reflection2}
\eea
where $R(u)\in End(V\otimes V)$, with $V$ being 8-dimensional
representation of $U_q[gl(3|1)]$, is the R-matrix of the two-parameter 
eight-state 
supersymmetric fermion model \cite{Gou97}, and $R_{21}(u)=P_{12}R_{12}(u)P_{12}$
with $P$ being the graded permutation operator;
the supertransposition $st_{\mu}~(\mu =1,2)$ 
is only carried out in the
$\mu$-th factor superspace of $V \otimes V$, whereas $ist_{\mu}$ denotes
the inverse operation of  $st_{\mu}$. 

The whole procedure of solving the reflection equations is quite
involved. We shall not spell out the details, but state that
there are three different diagonal boundary 
K-matrices, $K^I_-(u),~ K^{II}_-(u),~K^{III}_-(u)$, which solve the
first reflection equation (\ref{reflection1}):
\bea
K^I_-(u)&=&\frac {1}{\sinh\frac{ \xi^I_-}{2}\k\sinh(1+\frac{\xi^I_-}{2})\k
\sinh(2+\frac{\xi^I_-}{2})\k} {\rm diag}\lt ( 
  A^I_-(u),B^I_-(u),B^I_-(u),B^I_-(u),C^I_-(u),C^I_-(u),
  C^I_-(u),D^I_-(u)\rt),\no\\
K^{II}_-(u)&=&
\frac {1}
{\sinh\frac
{\xi^{II}_-}
{2}
\k\sinh(1+\frac
{\xi^{II}_-}
{2})
  \k }
  {\rm diag}\lt ( 
  A^{II}_-(u),A^{II}_-(u),B^{II}_-(u),B^{II}_-(u),B^{II}_-(u),B^{II}_-(u),
  C^{II}_-(u),C^{II}_-(u)\rt),\no\\
K^{III}_-(u)&=&\frac {1}{\sinh\frac{ \xi^{III}_-}{2}\k} {\rm diag}\lt ( 
  A^{III}_-(u),A^{III}_-(u),A^{III}_-(u),B^{III}_-(u),A^{III}_-(u),B^{III}_-(u),
  B^{III}_-(u),B^{III}_-(u)\rt),
\eea
where
\bea
A^I_-(u)&=&-e^{\frac{3}{2}u\k}\sinh\frac{-\xi^I_-+u}{2}\k\sinh
\frac{-2-\xi^I_-+u}{2}\k\sinh\frac{-4-\xi^I_- +u}{2}\k,\no\\
B^I_-(u)&=&e^{\frac{1}{2}u\k}\sinh\frac{\xi^I_-+u}{2}\k\sinh
\frac{-2-\xi^I_-+u}{2}\k\sinh\frac{-4-\xi^I_- +u}{2}\k,\no\\
C^I_-(u)&=&-e^{-\frac{1}{2}u\k}\sinh\frac{\xi^I_-+u}{2}\k\sinh
\frac{2+\xi^I_-+u}{2}\k\sinh\frac{-4-\xi^I_- +u}{2}\k,\no\\
D^I_-(u)&=&e^{-\frac{3}{2}u\k}\sinh\frac{\xi^I_-+u}{2}\k\sinh
\frac{2+\xi^I_-+u}{2}\k\sinh\frac{4+\xi^I_- +u}{2}\k,\no\\
A^{II}_-(u)&=&e^{u\k}\sinh\frac{-\xi^{II}_-+u}{2}\k\sinh\frac{-2-\xi^{II}_-+u}
{2}\k,\no\\
B^{II}_-(u)&=&-\sinh\frac{\xi^{II}_-+u}{2}\k\sinh\frac{-2-\xi^{II}_-+u}
{2}\k,\no\\
C^{II}_-(u)&=&e^{-u\k}\sinh\frac{\xi^{II}_-+u}{2}\k\sinh\frac{2+\xi^{II}_-+u}
{2}\k,\no\\
A^{III}_-(u)&=&-e^{\frac{u}{2}\k}\sinh\frac{-\xi^{III}_-+u}{2}\k,~~~~~~
  B^{III}_-(u)=e^{-\frac{u}{2}\k}\sinh\frac{\xi^{III}_-+u}{2}\k.
\eea
The corresponding K-matrices, $K^I_+(u),~K^{II}_+(u),~K^{III}_+(u)$,
can be obtained from the isomorphism of the
two reflection equations. Indeed, given a solution
$K^a_- (u)$ of (\ref{reflection1}), then $K^a_+(u)$ defined by
\beq
{K^a_+}^{st}(u) = M K^a_-(-u+2),~~~~~a=I,\,II,\;III,\label{t+t-}
\eeq
are solutions of (\ref{reflection2}). 
The proof follows from some algebraic computations upon
substituting (\ref{t+t-}) into  
(\ref{reflection2}) and making use
of the properties of the R-matrix . Here $M$ is the so-called
crossing matrix, which is given by in the present case,
\beq
M= {\rm diag} \lt (1,1,e^{2\k},e^{4\k},e^{2\k},e^{4\k},
e^{6\k},e^{6\k} \rt )
\eeq
Therefore, one may choose the boundary matrices $K^a_+(u)$ as 
\bea
K^I_+(u)&=&
  {\rm diag}\lt ( 
  A^I_+(u),B^I_+(u),e^{2\k}B^I_+(u),e^{4\k}B^I_+(u),C^I_+(u),e^{2\k}C^I_+(u),
 e^{4\k} C^I_+(u),D^I_+(u)\rt),\no\\
K^{II}_+(u)&=& {\rm diag}\lt ( 
  A^{II}_+(u),A^{II}_+(u),B^{II}_+(u),C^{II}_+(u),B^{II}_+(u),C^{II}_+(u),
  D^{II}_+(u),D^{II}_+(u)\rt),\no\\
K^{III}_+(u)&=& {\rm diag}\lt ( 
  A^{III}_+(u),A^{III}_+(u),e^{2\k}A^{III}_+(u),B^{III}_+(u),
  e^{2\k}A^{III}_+(u),B^{III}_+(u),
  e^{2\k}B^{III}_+(u),e^{2\k}B^{III}_+(u)\rt),
\eea
where
\bea
A^I_+(u)&=&e^{-\frac{3}{2}u\k}\sinh\frac{2g -\xi^I_++u}{2}\k\sinh\frac
{2g-2-\xi^I_++u}{2}\k\sinh\frac{2g-4-\xi^I _+ +u}{2}\k,\no\\
B^I_+(u)&=&e^{-(\frac{1}{2}u+2)\k}\sinh\frac{2g -\xi^I_+-u}{2}\k\sinh\frac
{2g-2-\xi^I_++u}{2}\k\sinh\frac{2g-\xi^I _+ +u}{2}\k,\no\\
C^I_+(u)&=&e^{(\frac{1}{2}u-2)\k}\sinh\frac{2g -\xi^I_+-u}{2}\k\sinh\frac
{2g+2-\xi^I_+-u}{2}\k\sinh\frac{2g-\xi^I _+ +u}{2}\k,\no\\
D^I_+(u)&=&e^{\frac{3}{2}u\k}\sinh\frac{2g -\xi^I_+-u}{2}\k\sinh\frac
{2g+2-\xi^I_+-u}{2}\k\sinh\frac{2g+4-\xi^I _+ -u}{2}\k,\no\\
A^{II}_+(u)&=&e^{-u\k}\sinh\frac{2g -\xi^{II}_++u}{2}\k\sinh\frac
{2g-2-\xi^{II}_++u}{2}\k,\no\\
B^{II}_+(u)&=&e^{\k}\sinh\frac
{2g-2-\xi^{II}_++u}{2}\k\sinh\frac{2g+2-\xi^{II} _+ -u}{2}\k,\no\\
C^{II}_+(u)&=&e^{2\k}\sinh\frac{2g -\xi^{II}_++u}{2}\k\sinh\frac
{2g+2-\xi^{II}_+-u}{2}\k,\no\\
D^{II}_+(u)&=&e^{(u+2)\k}\sinh\frac
{2g+2-\xi^{II}_+-u}{2}\k\sinh\frac{2g+4-\xi^{II} _+ -u}{2}\k,\no\\
A^{III}_+(u)&=&e^{-\frac{1}{2}u\k}
\sinh\frac{2g-\xi^{III} _+ +u}{2}\k,\no\\
B^{III}_+(u)&=&e^{(\frac{1}{2}u+2)\k}\sinh\frac{2g+4 -\xi^{III}_+-u}{2}\k
,\no\\
\eea

Following  Sklyanin's arguments \cite{Skl88}, one
may show that the quantity ${\cal T}_-(u)$ given by
\beq
{\cal T}_-(u) = T(u) {K}_-(u) T^{-1}(-u),~~~~~~ 
T(u) = R_{0L}(u)\cdots R_{01}(u),
\eeq
satisfies the same relation as $K_-(u)$:
\beq
R_{12}(u_1-u_2)\stackrel {1}{\cal T}_-(u_1) R_{21}(u_1+u_2)
  \stackrel {2}{\cal T}_-(u_2)
=  \stackrel {2}{\cal T}_-(u_2) R_{12}(u_1+u_2)
  \stackrel {1}{\cal T}_-(u_1) R_{21}(u_1-u_2).
\eeq
Thus if one defines the boundary transfer matrix $\tau(u)$ as
\beq
\tau(u) = str (K_+(u){\cal T}_-(u))=str\lt(K_+(u)T(u)K_-(u)T^{-1}(-u)\rt),
\eeq
then it can be shown \cite{Bra97} that
$[\tau(u_1),\tau(u_2)] = 0$. Since $K_\pm(u)$ can be taken as 
$K^I_\pm(u),~K^{II}_\pm(u)$ and $K^{III}_\pm(u)$, respectively, we have nine
possible choices of the boundary transfer matrices:
\beq
\tau^{(a,b)}(u)=str\lt(K^a_+(u)T(u)K^b_-(u)T^{-1}(-u)\rt),~~~~
  a,\;b=I,\;II,\;III,\label{t-matrices}
\eeq
which reflects the fact that the boundary conditions on the left end
and on the right end of the open lattice chain are independent. 

Now it can be shown  that 
Hamiltonians corresponding to all nine boundary conditions are related
to the second derivative of
the boundary transfer matrix
$\tau^{(a,b)} (u)$ (up to an unimportant additive constant)
\bea
H&=&\frac{2\sinh\k g}{\k}
   \;H^{(a,b)},\no\\ 
H^{(a,b)}&=&\frac {\tau^{(a,b)''} (0)}{4(V+2W)}=
  \sum _{j=1}^{L-1} H^R_{j,j+1} + \frac {1}{2} \stackrel {1}{K^{b'}}_-(0)
+\frac {1}{2(V+2W)}\lt[str_0\lt(\stackrel {0}{K^a}_+(0)G_{L0}\rt)\rt.\no\\
& &\lt.+2\,str_0\lt(\stackrel {0}{K^{a'}}_+(0)H_{L0}^R\rt)+
  str_0\lt(\stackrel {0}{K^a}_+(0)\lt(H^R_{L0}\rt)^2\rt)\rt],\label{derived-h}
\eea
where 
\bea
V&=&str_0 K^{a'}_+(0),
~~W=str_0 \lt(\stackrel {0}{K^a}_+(0) H_{L0}^R\rt),\no\\
H^R_{j,j+1}&=&P_{j,j+1}R'_{j,j+1}(0),
~~G_{j.j+1}=P_{j,j+1}R''_{j,j+1}(0).
\eea
Here $P_{j,j+1}$ denotes the graded permutation operator acting on the $j$-th
and $j+1$-th quantum spaces. (\ref{derived-h}) implies that the boundary
two-parameter eight-state supersymmetric models admit an infinite number
of conservation currents which are in involution with each other, thus
assuring their quantum integrability.

\sect{Coordinate Bethe Ansatz Analysis}

Having established the quantum integrability of the boundary models,
we now  solve them by using
the coordinate space Bethe ansatz method.
Following \cite{Asa96,Zha97,Bra97},
we assume that the eigenfunction of Hamiltonian (\ref{hamiltonian}) 
takes the form
\bea
| \Psi \rangle& =&\sum _{\{(x_j,\a_j)\}}\Psi _{\a_1,\cdots,\a_N}
  (x_1,\cdots,x_N)c^\dagger
  _{x_1\a_1}\cdots c^\dagger_{x_N\a_N} | 0 \rangle,\no\\
\Psi_{\a_1,\cdots,\a_ N}(x_1,\cdots,x_N)
&=&\sum _P \e _P A_{\a_{Q1},\cdots,\a_{QN}}(k_{PQ1},\cdots,k_{PQN})
\exp (i\sum ^N_{j=1} k_{P_j}x_j),
\eea
where the summation is taken over all permutations and negations of
$k_1,\cdots,k_N,$ and $Q$ is the permutation of the $N$ particles such that
$1\leq   x_{Q1}\leq   \cdots  \leq  x_{QN}\leq   L$.
The symbol $\e_P$ is a sign factor $\pm1$ and changes its sign
under each 'mutation'. Substituting the wavefunction into  the
eigenvalue equation $ H| \Psi  \rangle = E | \Psi \rangle $,
one gets
\bea
A_{\cdots,\a_j,\a_i,\cdots}(\cdots,k_j,k_i,\cdots)&=&S_{ij}(
    k_i,k_j)
    A_{\cdots,\a_i,\a_j,\cdots}(\cdots,k_i,k_j,\cdots),\no\\
A_{\a_i,\cdots}(-k_j,\cdots)&=&s^L(k_j;p_{1\a_i})A_{\a_i,\cdots}
    (k_j,\cdots),\no\\
A_{\cdots,\a_i}(\cdots,-k_j)&=&s^R(k_j;p_{L\a_i})A_{\cdots,\a_i}(\cdots,k_j),
\eea
where $S_{ij}(k_i,k_j) $ are
the two-particle scattering matrices,
\bea
S_{ij}(k_i,k_j)^{aa}_{aa}&=&1,~~~ a=1,2,3,\no\\
S_{ij}(k_i,k_j)^{ab}_{ab}&=&\frac {\sin (\l_i-\l_j)}
  {\sin (\l_i-\l_j-i\kappa)},~~~
  a \neq b,~a,b=1,2,3,\no\\
S_{ij}(k_i,k_j)^{ab}_{ba}&=&e^{i {\rm sign} (a-b)(\l_i-\l_j)}
\frac {\sin i\kappa}{\sin (\l_i-\l_j-i\kappa)},~~~ a,b=1,2,3,
\eea
where $\l_j$ are suitable particle
rapidities related to the quasi-momenta $k_j$ of the electrons by
\beq
k(\l)=
2 \arctan (\coth c \tan \l),
\eeq
where the parameter $c$ is defined by
\beq
c= \frac {1}{4} \lt\{ \ln [\frac {\sinh \frac {1}{2}(\eta +\kappa)}
 {\sinh \frac {1}{2}(\eta -\kappa)}]-\kappa \rt\}.
\eeq
 $~s^L(k_j;p_{1\a_i})$ and  
 $~s^R(k_j;p_{L\a_i})$ are the boundary scattering matrices,
\bea
s^L(k_j;p_{1\a_i})&=&\frac {1-p_{1\a_i}e^{ik_j}}
{1-p_{1\a_i}e^{-ik_j}},\no\\
s^R(k_j;p_{L\a_i})&=&\frac {1-p_{L\a_i}e^{-ik_j}}
{1-p_{L\a_i}e^{ik_j}}e^{2ik_j(L+1)}
\eea
with $p_{1\a_i}$ and $p_{L\a_i}$ being given by the following formulae,
corresponding to the nine cases, respectively,
\bea
{\rm Case~i:}~~
&&p_{1,1}=p_{1,2}=p_{1,3}\equiv p_1=
   -e^{-\k g}+ e^{-\frac{\xi^I_-}{2}\k}
   \frac{\sinh\k g}{\sinh\frac{\xi^I_-}{2}\k},\no\\
&&p_{L,1}=p_{L,2}=p_{L,3}\equiv p_L=
   -e^{-\k g}+e^{\frac{\xi^I_+}{2}\k}
   \frac{\sinh\k g}{\sinh\frac{\xi^I_+}{2}\k}
 ;\label {p1}\\
{\rm Case~ii:}~~
&&p_{1,1}=-e^{-\k g},~~~p_{1,2}=p_{1,3}=
   -e^{-\k g}+ e^{-\frac{\xi^{II}_-}{2}\k}
   \frac{\sinh\k g}{\sinh\frac{\xi^{II}_-}{2}\k},\no\\
&&p_{L,1}=-e^{-\k g},~~~p_{L,2}=p_{L,3}=
   -e^{-\k g}+ e^{\frac{\xi^{II}_+}{2}\k}
   \frac{\sinh\k g}{\sinh\frac{\xi^{II}_+}{2}\k}
  ;\label {p2}\\
{\rm Case~iii:}~~
&&p_{1,1}=p_{1,2}=-e^{-\k g},~~~p_{1,3}=
   -e^{-\k g}+e^{-\frac{\xi^{III}_-}{2}\k}
   \frac{\sinh\k g}{\sinh\frac{\xi^{III}_-}{2}\k},\no\\
&&p_{L,1}=p_{L,2}=-e^{-\k g},~~~p_{L,3}=
   -e^{-\k g}+ e^{\frac{\xi^{III}_+}{2}\k}
   \frac{\sinh\k g}{\sinh\frac{\xi^{III}_+}{2}\k}
  ;\label {p3}\\
{\rm Case~iv:}~~
&&p_{1,1}=p_{1,2}=p_{1,3}\equiv p_1=
   -e^{-\k g}+e^{-\frac{\xi^I_-}{2}\k}
   \frac{\sinh\k g}{\sinh\frac{\xi^I_-}{2}\k},\no\\
&&p_{L,1}=-e^{-\k g},~~~p_{L,2}=p_{L,3}=
   -e^{-\k g}+ e^{\frac{\xi^{II}_+}{2}\k}
   \frac{\sinh\k g}{\sinh\frac{\xi^{II}_+}{2}\k}
  ;\label {p4}\\
{\rm Case~v:}~~
&&p_{1,1}=-e^{-\k g},~~~p_{1,2}=p_{1-}=
   -e^{-\k g}+ e^{-\frac{\xi^{II}_-}{2}\k}
   \frac{\sinh\k g}{\sinh\frac{\xi^{II}_-}{2}\k},\no\\
&&p_{L,1}=p_{L,2}=p_{L,3}\equiv p_L=
   -e^{-\k g}+ e^{\frac{\xi^I_+}{2}\k}
   \frac{\sinh\k g}{\sinh\frac{\xi^I_+}{2}\k}
 ;\label {p5}\\
{\rm Case~vi:}~~
&&p_{1,1}=p_{1,2}=p_{1,3}\equiv p_1=
   -e^{-\k g}+ e^{-\frac{\xi^I_-}{2}\k}
   \frac{\sinh\k g}{\sinh\frac{\xi^I_-}{2}\k},\no\\
&&p_{L,1}=p_{L,2}=-e^{-\k g},~~~p_{L,3}=
   -e^{-\k g}+ e^{\frac{\xi^{III}_+}{2}\k}
   \frac{\sinh\k g}{\sinh\frac{\xi^{III}_+}{2}\k}
  ;\label {p6}\\
{\rm Case~vii:}~~
&&p_{1,1}=p_{1,2}=-e^{-\k g},~~~p_{1,3}=
   -e^{-\k g}+ e^{-\frac{\xi^{III}_-}{2}\k}
   \frac{\sinh\k g}{\sinh\frac{\xi^{III}_-}{2}\k},\no\\
&&p_{L,1}=p_{L,2}=p_{L,3}\equiv p_L=
   -e^{-\k g}+ e^{\frac{\xi^I_+}{2}\k}
   \frac{\sinh\k g}{\sinh\frac{\xi^I_+}{2}\k}
  ;\label {p7}\\
{\rm Case~viii:}~~
&&p_{1,1}=-e^{-\k g},~~~p_{1,2}=p_{1,3}=
   -e^{-\k g}+ e^{-\frac{\xi^{II}_-}{2}\k}
   \frac{\sinh\k g}{\sinh\frac{\xi^{II}_-}{2}\k},\no\\
&&p_{L,1}=p_{L,2}=-e^{-\k g},~~~p_{L,3}=
   -e^{-\k g}+ e^{\frac{\xi^{III}_+}{2}\k}
   \frac{\sinh\k g}{\sinh\frac{\xi^{III}_+}{2}\k}
  ;\label {p8}\\
{\rm Case~ix:}~~
&&p_{1,1}=p_{1,2}=-e^{-\k g},~~~p_{1,3}=
   -e^{-\k g}+ e^{-\frac{\xi^{III}_-}{2}\k}
   \frac{\sinh\k g}{\sinh\frac{\xi^{III}_-}{2}\k},\no\\
&&p_{L,1}=-e^{-\k g},~~~p_{L,2}=p_{L,3}=
   -e^{-\k g}+ e^{\frac{\xi^{II}_+}{2}\k}
   \frac{\sinh\k g}{\sinh\frac{\xi^{II}_+}{2}\k}
  .\label {p9}
\eea

As is seen above, the two-particle S-matrix (IV.3) is 
nothing but the R-matrix of the $U_q[gl(3)]$-invariant Heisenberg 
magnetic chain
and thus satisfies the quantum Yang-Baxter equation (QYBE),
\beq
S_{ij}(k_i,k_j)S_{il}(k_i,k_l)S_{jl}(k_j,k_l)=
S_{jl}(k_j,k_l)S_{il}(k_i,k_l)S_{ij}(k_i,k_j).
\eeq
It can be checked that the boundary scattering matrices $s^L$ and $s^R$ 
obey the reflection equations:
\bea
&&S_{ji}(-k_j,-k_i)s^L(k_j;p_{1\a_j})S_{ij}(-k_i,k_j)s^L(k_i;p_{1\a
  _i})\no\\
&&~~~~~~~~~~~~~~~~~~=s^L(k_i;p_{1\a _i})S_{ji}(-k_j,k_i)s^L(k_j;p_{1\a _i})
  S_{ij}(k_i,k_j),\no\\
&&S_{ji}(-k_j,-k_i)s^R(k_j;p_{L\a_j})S_{ij}(k_i,-k_j)s^R(k_i;p_{L\a
  _i})\no\\
&&~~~~~~~~~~~~~~~~~~= s^R(k_i;p_{L\a _i})S_{ji}(k_j,-k_i)s^R(k_j;p_{L\a _i})
  ;p_{\a_i})S_{ji}(k_j,k_i).\label{reflection-e}
\eea
This is seen as follows. One introduces the notation
\beq
s(k;p)=\frac  {1-pe^{-ik}}{1-p e^{ik}}.
\eeq
Then the boundary scattering matrices $s^L(k_j;p_{1\a_i})$,
 $~s^R(k_j;p_{L\a_i})$ can be written as, corresponding to the nine
cases, respectively,
\bea
{\rm Case~i:}~~&&s^L(k_j;p_{1\a_i})=s(-k_j;p_1)I,\no\\
&&s^R(k_j;p_{L\a_i})=e^{ik_j2(L+1)}s(k_j;p_L)I;\label{sa}\\
{\rm Case~ii:}~~&&s^L(k_j;p_{1\a_i})=s(-k_j;p_{1,1})\lt(
\begin{array}{ccc}
1 &0 &0\\
0&e^{2i\l_j}\frac{\sin(\zeta_-+\l_j)}
	       {\sin(\zeta_--\l_j)} & 0\\
0&0&e^{2i\l_j}\frac{\sin(\zeta_-+\l_j)}
	       {\sin(\zeta_--\l_j)}
\end{array}
\rt),\no\\
&&s^R(k_j;p_{L\a_i})=e^{ik_j2(L+1)}s(k_j;p_{L,1})\lt(
\begin{array}{ccc}
1 &0 &0\\
0&e^{2i\l_j}\frac{\sin(\zeta_+-\l_j)}
	       {\sin(\zeta_++\l_j)} & 0\\
0&0&e^{2i\l_j}\frac{\sin(\zeta_+-\l_j)}
	       {\sin(\zeta_++\l_j)}
\end{array}
\rt);\label{sb}\\
{\rm Case~iii:}~~&&s^L(k_j;p_{1\a_i})=
s(-k_j;p_{1,1})\lt(
\begin{array}{ccc}
1 &0 &0\\
0&1   & 0\\
0&0&e^{2i\l_j}\frac{\sin(\zeta'_-+\l_j)}
	       {\sin(\zeta'_--\l_j)}
\end{array}
\rt),\no\\
&&s^R(k_j;p_{L\a_i})=e^{ik_j2(L+1)}s(k_j;p_{L,1})\lt(
\begin{array}{ccc}
1&0&0\\
0&1&0\\
0&0&e^{2i\l_j}\frac{\sin(\zeta'_+-\l_j)}
	       {\sin(\zeta'_++\l_j)} 
\end{array}
\rt);\label{sc}\\
{\rm Case~iv:}~~&&s^L(k_j;p_{1\a_i})=
s(-k_j;p_1)I,\no\\
&&s^R(k_j;p_{L\a_i})=e^{ik_j2(L+1)}s(k_j;p_{L,1})\lt(
\begin{array}{ccc}
1 &0 &0\\
0&e^{2i\l_j}\frac{\sin(\zeta_+-\l_j)}
	       {\sin(\zeta_++\l_j)} & 0\\
0&0&e^{2i\l_j}\frac{\sin(\zeta_+-\l_j)}
	       {\sin(\zeta_++\l_j)}
\end{array}
\rt);\label{sd}\\
{\rm Case~v:}~~&&s^L(k_j;p_{1\a_i})=s(-k_j;p_{1,1})\lt(
\begin{array}{ccc}
1 &0 &0\\
0&e^{2i\l_j}\frac{\sin(\zeta_-+\l_j)}
	       {\sin(\zeta_--\l_j)} & 0\\
0&0&e^{2i\l_j}\frac{\sin(\zeta_-+\l_j)}
	       {\sin(\zeta_--\l_j)}
\end{array}
\rt),\no\\
&&s^R(k_j;p_{L\a_i})=e^{ik_j2(L+1)}s(k_j;p_L)I;\label{se}\\
{\rm Case~vi:}~~&&s^L(k_j;p_{1\a_i})=s(-k_j;p_1)I,\no\\
&&s^R(k_j;p_{L\a_i})=e^{ik_j2(L+1)}s(k_j;p_{L,1})\lt(
\begin{array}{ccc}
1&0&0\\
0&1&0\\
0&0&e^{2i\l_j}\frac{\sin(\zeta'_+-\l_j)}
	       {\sin(\zeta'_++\l_j)} 
\end{array}
\rt);\label{sf}\\
{\rm Case~vii:}~~&&s^L(k_j;p_{1\a_i})=
s(-k_j;p_{1,1})\lt(
\begin{array}{ccc}
1 &0 &0\\
0&1   & 0\\
0&0&e^{2i\l_j}\frac{\sin(\zeta'_-+\l_j)}
	       {\sin(\zeta'_--\l_j)}
\end{array}
\rt),\no\\
&&s^R(k_j;p_{L\a_i})=e^{ik_j2(L+1)}s(k_j;p_L)I;\label{sg}\\
{\rm Case~viii:}~~&&s^L(k_j;p_{1\a_i})=s(-k_j;p_{1,1})\lt(
\begin{array}{ccc}
1 &0 &0\\
0&e^{2i\l_j}\frac{\sin(\zeta_-+\l_j)}
	       {\sin(\zeta_--\l_j)} & 0\\
0&0&e^{2i\l_j}\frac{\sin(\zeta_-+\l_j)}
	       {\sin(\zeta_--\l_j)}
\end{array}
\rt),\no\\
&&s^R(k_j;p_{L\a_i})=e^{ik_j2(L+1)}s(k_j;p_{L,1})\lt(
\begin{array}{ccc}
1&0&0\\
0&1&0\\
0&0&e^{2i\l_j}\frac{\sin(\zeta'_+-\l_j)}
	       {\sin(\zeta'_++\l_j)} 
\end{array}
\rt);\label{sh}\\
{\rm Case~ix:}~~&&s^L(k_j;p_{1\a_i})=
s(-k_j;p_{1,1})\lt(
\begin{array}{ccc}
1 &0 &0\\
0&1   & 0\\
0&0&e^{2i\l_j}\frac{\sin(\zeta'_-+\l_j)}
	       {\sin(\zeta'_--\l_j)}
\end{array}
\rt),\no\\
&&s^R(k_j;p_{L\a_i})=e^{ik_j2(L+1)}s(k_j;p_{L.1})\lt(
\begin{array}{ccc}
1 &0 &0\\
0&e^{2i\l_j}\frac{\sin(\zeta_+-\l_j)}
	       {\sin(\zeta_++\l_j)} & 0\\
0&0&e^{2i\l_j}\frac{\sin(\zeta_+-\l_j)}
	       {\sin(\zeta_++\l_j)}
\end{array}
\rt);\label{si}
\eea
Here $I$ stands for $3\times 3$ identity matrix and 
$p_{1+},~p_{L+}$ are the ones given in (\ref{p2});$\zeta _{\pm},
\zeta' _{\pm}$ are parameters defined by
\beq
\zeta _{\pm}=\frac {g-\xi^{II}_{\pm}}{2i}\k,~~~~
\zeta'_{\pm}=\frac {g-\xi^{III}_{\pm}}{2i}\k.
\eeq
We immediately see that (\ref{sa}) are the trivial solutions of the reflection
equations (\ref{reflection-e}), whereas (\ref{sb}) and (\ref{sc})
are the diagonal
solutions \cite{Skl88,Mez91}. 

The diagonalization of Hamiltonian (\ref{hamiltonian}) reduces 
to solving  the following matrix  eigenvalue equation
\beq
T_jt= t,~~~~~~~j=1,\cdots,N,
\eeq
where $t$ denotes an eigenvector on the space of the spin variables
and $T_j$ takes the form
\beq
T_j=S_j^-(k_j)s^L(-k_j;p_{1\s_j})R^-_j(k_j)R^+_j(k_j)
    s^R(k_j;p_{L\s_j})S^+_j(k_j)
\eeq
with
\bea
S_j^+(k_j)&=&S_{j,N}(k_j,k_N) \cdots S_{j,j+1}(k_j,k_{j+1}),\no\\
S^-_j(k_j)&=&S_{j,j-1}(k_j,k_{j-1})\cdots S_{j,1}(k_j,k_1),\no\\
R^-_j(k_j)&=&S_{1,j}(k_1,-k_j)\cdots S_{j-1,j}(k_{j-1},-k_j),\no\\
R^+_j(k_j)&=&S_{j+1,j}(k_{j+1},-k_j)\cdots S_{N,j}(k_N,-k_j).
\eea
This problem can  be solved using the algebraic Bethe ansatz method.
The Bethe ansatz equations for all the nine cases are 
\bea
e^{ik_j2(L+1)}F(k_j;p_{1+},p_{L+})
&=&\prod_{\s=1}^{M_1}\frac{\sin (\l_j-\L^{(1)}_\s+i\kappa/2)}
      {\sin (\l_j-\L^{(1)}_\s-i\kappa/2)}
     \frac{\sin (\l_j+\L^{(1)}_\s+i\kappa/2)}
      {\sin (\l_j+\L^{(1)}_\s-i\kappa/2)},\no\\
\prod_{j=1}^N\frac{\sin (\L^{(1)}_\s-\l_j+i\kappa/2)}{\sin 
(\L^{(1)}_\s-\l_j-i\kappa/2)}
\frac {\sin (\L^{(1)}_\s+\l_j+i\kappa/2)}{\sin 
(\L^{(1)}_\s+\l_j-i\kappa/2)}&=&
   G(\L^{(1)}_\s;\zeta_-,\zeta_+)\prod_{\stackrel {\rho=1}{\rho \neq \s}}^{M_1}
   \frac{\sin (\L^{(1)}_\s-\L^{(1)}_\rho+i\kappa)}
   {\sin (\L^{(1)}_\s-\L^{(1)}_\rho-i\kappa)}
   \frac{\sin (\L^{(1)}_\s-\L^{(1)}_\rho+i\kappa)}
   {\sin (\L^{(1)}_\s-\L^{(1)}_\rho-i\kappa)}\no\\
\prod_{\rho=1}^{M_2}&&\frac{\sin (\L^{(1)}_\s-\L^{(2)}_\rho-i\kappa/2)}
   {\sin (\L^{(1)}_\s-\L^{(2)}_\rho+i\kappa/2)}
   \frac{\sin (\L^{(1)}_\s+\L^{(2)}_\rho-i\kappa/2)}
   {\sin (\L^{(1)}_\s+\L^{(2)}_\rho+i\kappa/2)},~~~\s=1,\cdots,M_1, \no\\
\prod_{\rho=1}^{M_1}\frac{\sin (\L^{(2)}_\g-\L^{(1)}_\rho+i\kappa/2)}
   {\sin (\L^{(2)}_\g-\L^{(1)}_\rho-i\kappa/2)}
\frac{\sin (\L^{(2)}_\g+\L^{(1)}_\rho+i\kappa/2)}
  {\sin (\L^{(2)}_-\g+L^{(1)}_\rho-i\kappa/2)}&=&K(\L^{(2)};\zeta'_-,\zeta'_+)
   \prod_{\stackrel {\rho=1}{\rho \neq \g}}^{M_2}
   \frac{\sin (\L^{(2)}_\g-\L^{(2)}_\rho+i\kappa)}
   {\sin (\L^{(2)}_\g-\L^{(2)}_\rho-i\kappa)}
   \frac{\sin (\L^{(2)}_\g+\L^{(2)}_\rho+i\kappa)}
   {\sin (\L^{(2)}_\g+\L^{(2)}_\rho-i\kappa)},\no\\
   & &\g=1,\cdots,M_2,\label{Bethe-ansatz}
\eea
where 
\bea
F(k_j;p_{1,1},p_{L,1})&=&s(k_j;p_{1,1})s(k_j;p_{L,1}),~~~( {\rm for \;all
\;cases}),\no\\
G(\L^{(1)}_\s;\zeta_-,\zeta_+)&=&
 \left \{ \begin {array}{ll}
1 & case \;(i)\\
\frac {\sin (\zeta _-+\Lambda ^{(1)}_\s +\frac {i\k}{2})}
 {\sin(\zeta _--\Lambda ^{(1)}_\s +\frac {i\k}{2})}
\frac {\sin(\zeta _++\Lambda ^{(1)}_\s +\frac {i\k}{2})}
 {\sin(\zeta _+-\Lambda ^{(1)}_\s +\frac {i\k}{2})}
& case\;( ii)\\
1 & case \;(iii)\\
\frac {\sin(\zeta _++\Lambda ^{(1)}_\s +\frac {i\k}{2})}
 {\sin(\zeta _+-\Lambda ^{(1)}_\s +\frac {i\k}{2})}e^{2i\Lambda
 ^{(1)}_\s}
& case \;(iv)\\
\frac {\sin(\zeta _-+\Lambda ^{(1)}_\s +\frac {i\k}{2})}
 {\sin(\zeta _--\Lambda ^{(1)}_\s +\frac {i\k}{2})}e^{-2i\Lambda^{(1)}_\s}
& case \;(v)\\
1 & case \;(vi)\\
1 & case \;(vii)\\
\frac {\sin(\zeta _-+\Lambda ^{(1)}_\s +\frac {i\k}{2})}
 {\sin(\zeta _--\Lambda ^{(1)}_\s +\frac {i\k}{2})}e^{-2i\Lambda
 ^{(1)}_\s}
& case \;(viii)\\
\frac {\sin(\zeta _++\Lambda ^{(1)}_\s +\frac {i\k}{2})}
 {\sin(\zeta _+-\Lambda ^{(1)}_\s +\frac {i\k}{2})}e^{2i\Lambda^{(1)}_\s}
& case\;( ix)
\end {array}\right.\no\\
K(\L^{(2)};\zeta'_-,\zeta'_+)&=&
 \left \{ \begin {array}{ll}
1 & case\;( i)\\
1 & case \;(ii)\\
\frac {\sin(\zeta' _-+\Lambda ^{(2)}_\g +i\k)}
 {\sin(\zeta' _--\Lambda ^{(2)}_\g +i\k)}
\frac {\sin(\zeta' _++\Lambda ^{(2)}_\g +i\k)}
 {\sin(\zeta' _+-\Lambda ^{(2)}_\g +i\k)}
& case \;(iii)\\
1 & case \;(iv)\\
1 & case \;(v)\\
\frac {\sin(\zeta' _++\Lambda ^{(2)}_\g +i\k)}
 {\sin(\zeta' _+-\Lambda ^{(2)}_\g +i\k)}e^{2i\Lambda^{(2)}_\g}
& case \;(vi)\\
\frac {\sin(\zeta' _-+\Lambda ^{(2)}_\g +i\k)}
 {\sin(\zeta' _--\Lambda ^{(2)}_\g +i\k)}e^{-2i\Lambda^{(2)}_\g}
& case \;(vii)\\
\frac {\sin(\zeta' _++\Lambda ^{(2)}_\g +i\k)}
 {\sin(\zeta' _+-\Lambda ^{(2)}_\g +i\k)}e^{2i\Lambda^{(2)}_\g}
& case \;(viii)\\
\frac {\sin(\zeta' _-+\Lambda ^{(2)}_\g +i\k)}
 {\sin(\zeta' _--\Lambda ^{(2)}_\g +i\k)}e^{-2i\Lambda^{(2)}_\g}
& case \;(ix)
\end {array}\right.\no\\
\eea
The energy eigenvalue $E$ of the model is given by
$E=-2\sum ^N_{j=1}\cos k_j$ (modular an unimportant additive constant coming
from the chemical potential term).

\vskip.3in
\acknowledgments
X.-Y. Ge is supported by an Australian Overseas Postgraduate
Research Scholarship. Y.-Z. Zhang
is supported by the QEII Fellowship Grant from Australian Research 
Council. H.-Q. Zhou would like to thank the department of mathematics,
University of Queensland, for kind hospitality. He is supported by the
National Natural Science Foundation of China and Sichuan Young 
Investigators Science and Technology Fund.



\begin{thebibliography}{99}
\bibitem {Ess94} F.H.L. Essler, V.E. Korepin, {\it Exactly solvable
   models of strongly correlated electrons}, World Scientific, 1994.
\bibitem{Bar91} R.Z. Bariev, J. Phys. {\bf A:} Math. Gen. {\bf 24}
   (1991) L919;\\
   A. Gonz\'alez-Ruiz, Nucl. Phys. {\bf B424} (1994) 553;\\
   H.-Q. Zhou, Phys. Rev. {\bf B54} (1996) 41; ibid
   {\bf B53} (1996) 5089.
\bibitem{Bra95} A.J. Bracken, M.D. Gould, J.R. Links, Y.-Z. Zhang,
   Phys. Rev. Lett. {\bf 74} (1995) 2768.
\bibitem{Bar95} R.Z. Bariev, A. Kl\"umper, J. Zittartz, Europhys. Lett.
   {\bf 32} (1995) 85;\\
   M.D. Gould, K.E. Hibberd, J.R. Links, Y.-Z. Zhang,
   Phys. Lett. {\bf A212} (1996) 156.
\bibitem{Asa96} H. Asakawa, M. Suzuki, J. Phys. {\bf A:} Math. Gen. 
   {\bf 29} (1996) 225;\\
    M. Shiroishi, M. Wadati, J. Phys. Soc. Jpn. {\bf 66} (1997) 1.
\bibitem{Gou97} M.D. Gould, Y.-Z. Zhang, H.-Q. Zhou, Phys. Rev.
   {\bf B}, in press, {\tt cond-mat/9709129};\\
   X.Y. Ge, M.D. Gould, Y.-Z. Zhang, H.-Q. Zhou, preprint
   cond-mat/9711305.
\bibitem{Bra97} A.J. Bracken, X.Y. Ge, Y.-Z. Zhang, H.-Q. Zhou, 
   Nucl. Phys. {\bf B}, in press, {\tt cond-mat/9710141};\\
   Nucl. Phys. {\bf B}, in press, {\tt cond-mat/9710171}.
\bibitem{Skl88} E.K. Sklyanin, J. Phys. {\bf A:} Math.Gen. {\bf 21} (1988)
   2375.
\bibitem{Mez91} L. Mezincescu, R. Nepomechie, J. Phys. {\bf A:} Math.
   Gen. {\bf 24} (1991) L17;\\
   H.J. de Vega, A. Gonz\'alez-Ruiz, J. Phys. {\bf A:}
   Math. Gen. {\bf 26} (1993) L519. 
\bibitem{Zha97} Y.-Z. Zhang, H.-Q. Zhou, preprints cond-mat/9707263,
   cond-mat/9711238.

\end{thebibliography}
\end{document}